\documentclass[
reprint,
amsmath,amssymb, 
pra, 
jmp,
floatfix,
superscriptaddress,
]{revtex4-2}

\usepackage{graphicx}% Include figure files
\usepackage{dcolumn}% Align table columns on decimal point
\usepackage{bm}% bold math

\newcommand{\LBL}{Chemical Sciences Division, Lawrence Berkeley National Laboratory, Berkeley, California 94720, USA}
\newcommand{\UCBchem}{Department of Chemistry, University of California, Berkeley, California 94720, USA}
\newcommand{\UCBphys}{Department of Physics, University of California, Berkeley, California 94720, USA}

\begin{document}

%\preprint{}

\title{Extracting doubly-excited state lifetimes in helium directly in the time domain with attosecond noncollinear four-wave-mixing spectroscopy}% Force line breaks with \par

\author{Patrick Rupprecht}
\email{prupprecht@lbl.gov}
\altaffiliation{These authors contributed equally to this work.}
\affiliation{\UCBchem}
\affiliation{\LBL}
\author{Nicolette G. Puskar}
\email{nicolette.puskar@berkeley.edu}
\altaffiliation{These authors contributed equally to this work.}
\affiliation{\UCBchem}
\affiliation{\LBL}
\author{Daniel M. Neumark}
\email{dneumark@berkeley.edu}
\affiliation{\UCBchem}
\affiliation{\LBL}
\author{Stephen R. Leone}
\email{srl@berkeley.edu}
\affiliation{\UCBchem}
\affiliation{\LBL}
\affiliation{\UCBphys}

\date{\today}

\begin{abstract}
The helium atom, with one nucleus and two electrons, is a prototypical system to study quantum many-body dynamics. Doubly-excited states, or quantum states in which both electrons are excited by one photon, are showcase scenarios of electronic-correlation mediated effects. In this paper, the natural lifetimes of the doubly-excited $^1$P$^o$ 2s$n$p Rydberg series and the $^1$S$^e$ 2p$^2$ dark state in helium in the 60\,eV to 65\,eV region are measured directly in the time domain with extreme-ultraviolet/near-infrared noncollinear attosecond four-wave-mixing (FWM) spectroscopy. The measured lifetimes are in agreement with lifetimes deduced from spectral linewidths and theoretical predictions, and the roles of specific decay mechanisms are considered. While complex spectral line shapes in the form of Fano resonances are common in absorption spectroscopy of autoionizing states, the background-free and thus homodyned character of noncollinear FWM results exclusively in Lorentzian spectral features in the absence of strong-field effects. The onset of strong-field effects that would affect the extraction of accurate natural lifetimes in helium by FWM is determined to be approximately 0.3 Rabi cycles. This study provides a systematic understanding of the FWM parameters necessary to enable accurate lifetime extractions, which can be utilized in more complex quantum systems such as molecules in the future.
\end{abstract}

\maketitle

%\tableofcontents

\section{\label{sec:intro}Introduction}

Quantum dynamics governs the interaction of \mbox{(sub-)}atomic particles. 
In addition to the interaction between the electrons and the nucleus, the interaction between the electrons themselves plays a crucial role within intra-atomic dynamics on the shortest time scales \cite{rudenko2004correlated}.
A prime example of electronic-correlation related phenomena are doubly-excited states. 
Here, one photon excites two electrons at the same time, which is mediated by the interaction between the two electrons.
The simplest quantum mechanical system that realizes this special case of light-matter interaction is the helium atom with one nucleus and two electrons. 
For example, both 1s electrons can be excited by one 60.15\,eV extreme ultraviolet (XUV) photon to a 2s2p two-electron quantum state. 
In addition, other 2s$n$p states within the Rydberg series, which converges to the He$^+$ ($n$=2) ionization threshold around E$_{IP}$($n$=2)$ \ = 65.4$\,eV, can also be excited by XUV photons of higher energy.
These doubly-excited states in helium have drawn significant attention since they were first observed in electron-scattering \cite{whiddington1934note, silverman1964additional} and absorption measurements \cite{madden1963new} as well as theoretically described \cite{fano1961effects} and classified \cite{cooper1963classification}. 
As the 2s$n$p doubly-excited states in helium are located in the 60\,eV to 65\,eV spectral region, they are energetically located above the first ionization potential of helium at 24.59\,eV \cite{kandula2010extreme}, hence resulting in their metastable, autoionizing character with lifetimes spanning tens to hundreds of femtoseconds.
High-resolution XUV absorption spectroscopy at synchrotrons \cite{domke1991extensive, domke1992observation, domke1996high, schulz1996observation} was used to characterize the spectral widths $\Gamma$ of these resonances, which in the case of gas-phase atoms at low pressures directly translate to the natural lifetimes of the states by the relationship $T = \hbar / \Gamma$. %http://hyperphysics.phy-astr.gsu.edu/hbase/quantum/parlif.html
As the doubly-excited 2s$n$p states are embedded in the continuum, they form a quantum interferometer, in which the pathways of direct photoionization and autoionization via the metastable doubly-excited states interfere with each other \cite{ott2014reconstruction}. 
This gives rise to the Fano spectral line shapes of these resonances \cite{fano1961effects}.
In recent decades, the Fano character of the 2s$n$p states has been actively exploited to demonstrate quantum control within atoms using ultrashort laser pulses \cite{gilbertson2010monitoring,ott2013lorentz,gruson2016attosecond,kaldun2016observing}. \par
Among the current theoretical advances is a publication by Mi \textit{et al.} \cite{mi2021method} that suggests applying XUV/near-infrared (NIR) noncollinear attosecond four-wave-mixing (FWM) spectroscopy to measure the doubly-excited 2s$n$p state lifetimes in helium directly in the time domain and in a quantum-state specific manner.
Since its first demonstration in 2016 \cite{cao2016noncollinear}, XUV/NIR FWM spectroscopy has revealed detailed insights into the dynamics of gas-phase atoms \cite{fidler2019nonlinear, marroux2018multidimensional, fidler2019autoionization, puskar2023measuring, gaynor2023nonresonant}, molecules \cite{cao2018excited,lin2021coupled, fidler2022state}, and solids \cite{gaynor2021solid}.
The capability to access few-femtosecond natural lifetimes in a state-specific manner in the time domain is unique to noncollinear XUV/NIR FWM spectroscopy. 
Furthermore, XUV/NIR FWM as a nonlinear spectroscopic method grants access to states with dipole-allowed ('bright') and dipole-forbidden ('dark') transitions from the ground state while still relying on an all-optical measurement scheme. \par
So far, XUV/NIR FWM studies in helium concentrated on singly-excited states between 20\,eV and 25\,eV \cite{fidler2019nonlinear,fidler2020self}. 
These singly-excited Rydberg states have long lifetimes on the order of nanoseconds \cite{theodosiou1984lifetimes}.
Now there is renewed interest in measuring and understanding ultrafast decay mechanisms, especially with the availability of attosecond pulses \cite{krausz2009attosecond}.
Concerning doubly-excited states in helium, Gilbertson \textit{et al.} reported on the time-domain characterization of the 2s2p lifetime via attosecond streaking spectroscopy \cite{gilbertson2010monitoring}. 
To our knowledge, no time-domain lifetime measurements of additional 2s$n$p bright states or any doubly-excited dark states in helium have been reported so far.\par
In this paper, we conduct XUV/NIR FWM spectroscopy on the 2s$n$p bright-states in helium up to $n=6$ as well as the 2p$^2$ dark state and characterize their natural lifetimes for the first time directly in the time domain. 
In addition, the spectral line shapes of the FWM signal in the low-NIR-intensity limit are found to be exclusively Lorentzian in contrast to the Fano character of the static absorption spectra of autoionizing states.
For higher NIR intensities, the threshold for the onset of strong-field effects that alter the observed lifetimes is determined to be approximately 0.3 Rabi cycles. 

\section{\label{sec:exp}Experimental methods}

\begin{figure}[h]
\centering
\includegraphics[width=0.48\textwidth]{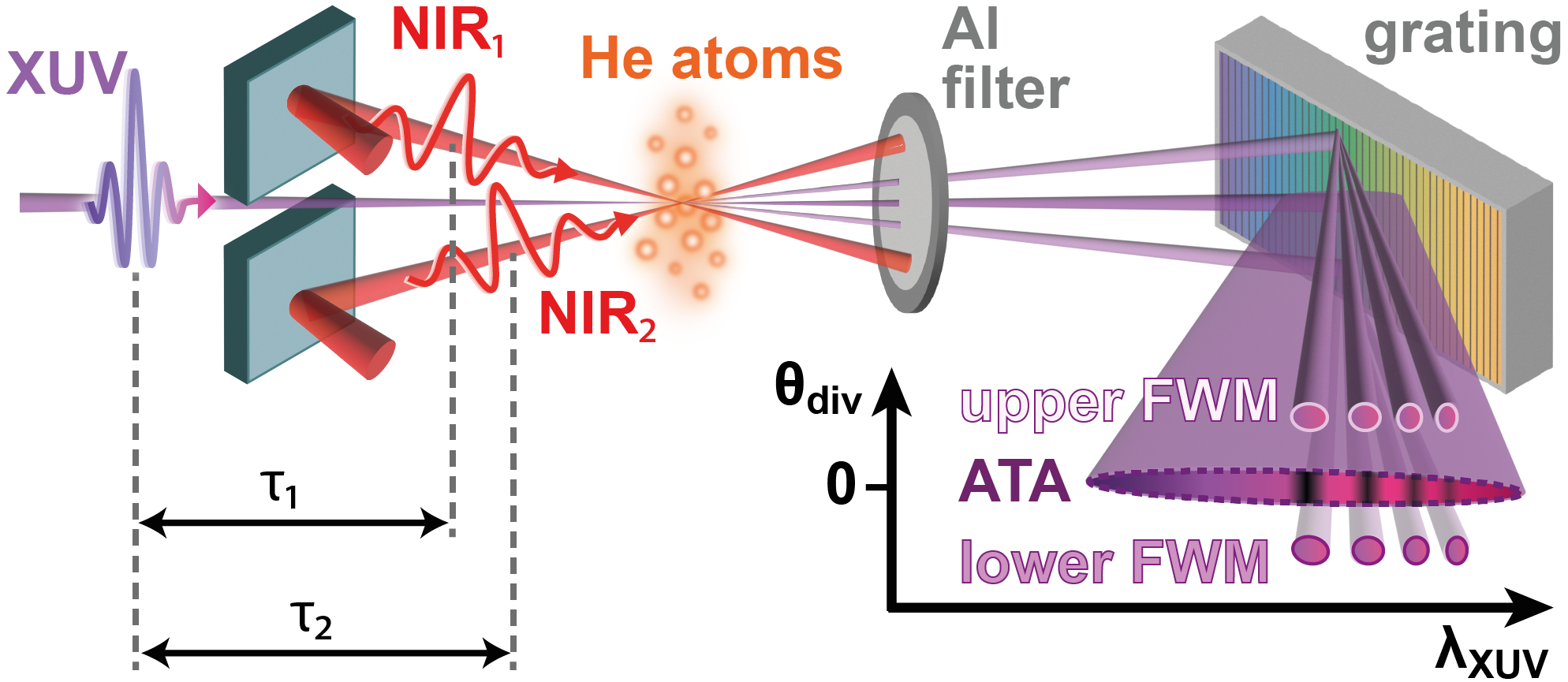}
\caption{Experimental scheme of the noncollinear XUV/NIR FWM measurement. A broadband, attosecond XUV pulse is focused into a gas-phase helium target together with two time-delayable few-cycle near infrared pulses (NIR$_1$ and NIR$_2$). The $\chi^{(3)}$ wave-mixing process of the two noncollinear NIR pulses with the XUV pulse results in the emission of FWM signatures that appear above and below the transmitted attosecond transient absorption (ATA) spectroscopy beam according to the phase-matching Eq.~\ref{eq:phase-matching}. The spectra of the spatially isolated XUV signals (divergence axis $\Theta_{div}$) are recorded by the XUV-sensitive camera chip for different time delays $\tau_1$ and $\tau_2$. }
\label{fig:expsetup}
\end{figure}

The experimental setup for noncollinear attosecond XUV/NIR FWM spectroscopy of doubly-excited states in helium is shown in Fig.~\ref{fig:expsetup}.

\noindent First, a 3.8\,mJ portion of the output of a commercial Titanium:Sapphire laser system (Coherent Legend Elite Duo HE+ USP; 1\,kHz repetition rate, 35\,fs pulse duration, 800\,nm center wavelength) is spectrally broadened in a stretched, neon-filled hollow core fiber (few-cycle Inc.).
The resulting supercontinuum is temporally compressed to few-cycle duration with chirped mirrors (Ultrafast Innovations PC70), glass wedges, and thin ADP crystals, for third-order-dispersion compensation \cite{timmers2017generating}. 
The majority (70\%) of the pulse energy is used to generate XUV attosecond pulses (a pulse train consisting of a few attosecond pulses) with broadband, continuous spectra in the 40\,eV to 70\,eV photon-energy region via high-order harmonic generation (HHG) in argon, while the rest (30\%) is sent into an interferometric assembly to establish the NIR wave-mixing pulses.
Here, the beam is further split into two beams, which are independently time delayed ($\tau_1$ and $\tau_2$ in Fig.~\ref{fig:expsetup}) with respect to the XUV pulse via piezoelectric-actuator-controlled optical delay stages.
Afterwards, the two few-cycle NIR beams (NIR$_1$ and NIR$_2$ in Fig.~\ref{fig:expsetup}) are focused (f = 1\,m) into a helium-filled (40\,mbar backing pressure) effusive gas cell (500\,$\mu$m through-hole diameter, 2\,mm inner tube diameter) within the vacuum beamline.
After HHG, residual NIR light is removed using a 200\,nm thick aluminum (Al) filter.
Moreover, the XUV pulses are refocused under grazing incidence ($\Theta_{grazing} = 10^\circ$) into the target cell using a gold-coated toroidal mirror.
The NIR few-cycle pulses propagate at an angle of approximately $\Theta_{NIR1,NIR2} = \pm 1.5^\circ$ with respect to the XUV pulse and all three beams---XUV, NIR$_1$, and NIR$_2$---are spatially overlapped in the target cell. \par
In a $\Lambda$- or V-type FWM coupling scheme, addition of an NIR$_1$ (or NIR$_2$) photon from an XUV-excited state and subtraction of an NIR$_2$ (or NIR$_1$) photon back to the same XUV-excited state leads to the emission of degenerate XUV four-wave-mixing signals.
Due to the noncollinear optical geometry, the generated FWM signals are emitted at a small angle with respect to the attosecond transient absorption (ATA) signal according to the phase-matching relation for the wave vectors
\begin{equation}
    \vec{k}_{FWM} = \vec{k}_{XUV} \mp \vec{k}_{NIR_1} \pm \vec{k}_{NIR_2}\, , 
\label{eq:phase-matching}
\end{equation}
with the signs in the equation corresponding to the upper and lower FWM signals, respectively, on the XUV spectrograph camera.
The noncollinear angles $\Theta_{NIR_1}$ and $\Theta_{NIR_2}$ of the input NIR beams are chosen such that the resulting FWM emission angles 
\begin{equation}
    \Theta_{FWM} \approx \pm \frac{\Theta_{NIR_1} \, E_{NIR_1} + \Theta_{NIR_2} \, E_{NIR_2}}{E_{XUV}}
\end{equation}
for an XUV-excited state with energy $E_{XUV}$ coupled to/from an intermediate state with NIR photons of energy $E_{NIR_1}$ and $E_{NIR_2}$ are larger than the divergence of the transmitted ATA XUV beam. 
\begin{figure*}[t]
\centering
\includegraphics[width=0.9\textwidth]{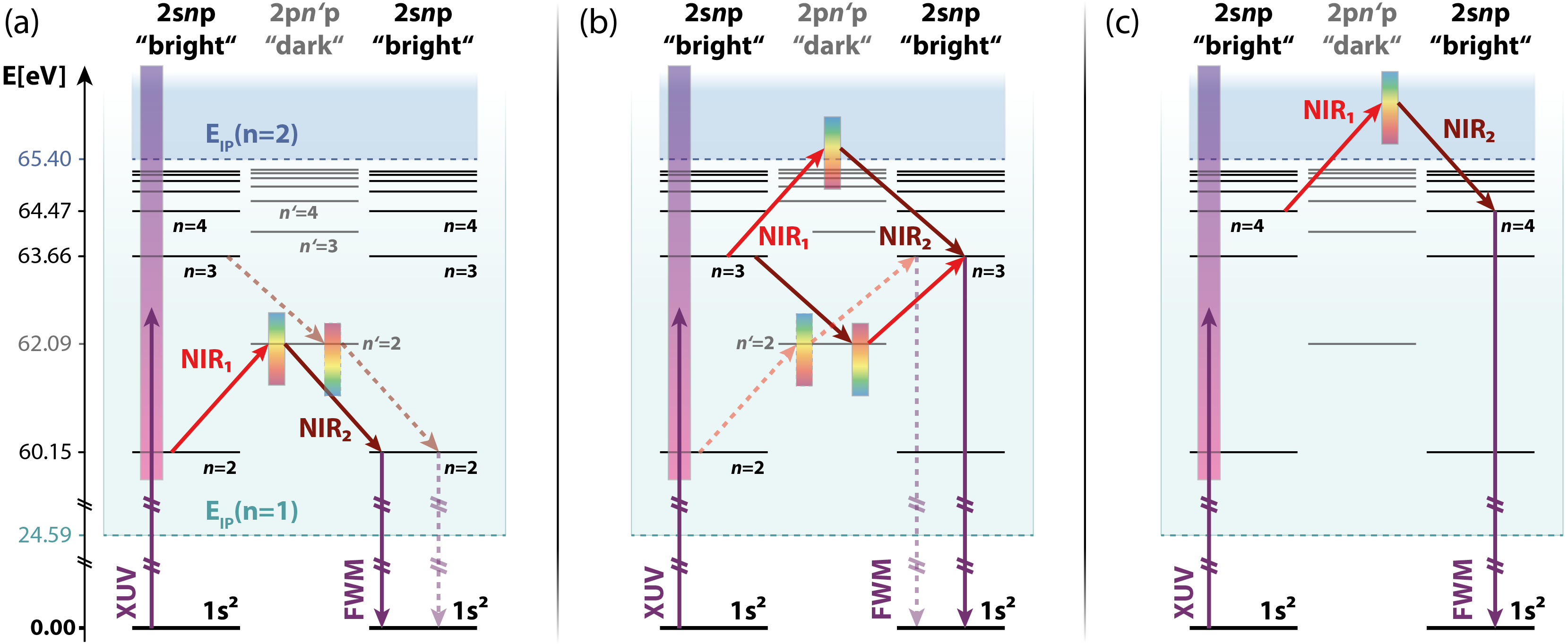}
\caption{FWM-coupling scheme of doubly excited states in helium. The broadband attosecond XUV pulse (purple vertical spectral bar) populates the 2s$n$p Rydberg series, which is embedded in the helium 1s ionization continuum (light blue) and converges to the 2s continuum (dark blue). The bright and dark red arrows indicate the FWM couplings via the NIR$_1$ and NIR$_2$ pulses, respectively, with the red-to-blue vertical spectral bars representing the energy region that can be resonantly coupled within the NIR few-cycle pulse spectrum. Only couplings are depicted that are relevant for the measured lower FWM signal on the XUV spectrograph camera in Fig.~\ref{fig:expsetup}. Three cases are differentiated: (a) The FWM signal at the 2s2p energy originates from a $\Lambda$-coupling to the 2p$^2$ dark state. While a ladder coupling of two NIR$_2$ pulses from the 2s3p state is in principle also possible (dashed arrows), it is much weaker due to the smaller transition dipole moment between the 2s3p and the 2p$^2$ states. (b) The FWM signature at the 2s3p energy consists of multiple $\Lambda$ and V-coupling pathways to discrete dark states as well as the 2s continuum. A potential ladder coupling via two NIR$_1$ photons from the 2s2p state will only influence the FWM signal close to the temporal overlap due to the significantly shorter 2s2p lifetime in comparison to the 2s3p lifetime. (c) The FWM signatures at the 2s$n$p Rydberg energies with $n \geq 4$ are due to $\Lambda$-couplings to the 2s continuum.}
\label{fig:fwmscheme}
\end{figure*}
After the target cell, the residual wave-mixing NIR pulses are removed from the three XUV signatures (ATA, upper and lower FWM in Fig.~\ref{fig:expsetup}) by employing another 200\,nm thick Al filter. 
The XUV beams are spectrally dispersed by a variable-line-spaced reflective grating (Hitachi 001-0640) and the resulting spectra are recorded using an XUV-sensitive CCD camera (Princeton Instruments PIXIS XO-400B).
In order to optimize the signal-to-noise ratio of the recorded FWM signal, the CCD camera is vertically moved to mask the intense ATA spectrum and only record the spatially lower FWM signal.\par
If the lifetimes of bright states are targeted, both NIR pulses remain in temporal overlap ($\tau_1 = \tau_2$), while their delay with respect to the XUV pulse is scanned. 
Such a scheme is called a \textit{bright-state scan}.
If the lifetime of the dark state is desired, the NIR$_1$ pulse remains in temporal overlap with the XUV pulse ($\tau_1 = 0$), while the delay of NIR$_2$ is varied.
Hence, this pulse sequence is called a \textit{dark-state scan}.
These two different schemes have been successfully employed in previous experiments, e.g., by Puskar \textit{et al.} to characterize inner-valence-excited state lifetimes of bright and dark states in neon atoms \cite{puskar2023measuring}.\par
The goal of measuring the natural lifetimes of various XUV-excited states makes a precision control of the NIR intensities in the target cell over multiple orders of magnitude mandatory to avoid effects such as Rabi cycling, six wave mixing, Stark effects, and pumping to higher excited states. 
Utilizing a variable aperture in the collimated NIR FWM beams for intensity control grants this flexibility without altering the pulse duration, as would occur by introducing additional dispersive elements like neutral density filters or polarizers and waveplates.
For low intensity measurements ($< 10^{12}$\,W/cm$^2$), an additional 80:20 beamsplitter is installed to avoid extremely large NIR foci due to diffraction from extensive aperture closure.
A thorough characterization of the resulting beam diameter and temporal structure enables translating the measured power after the closed aperture into a respective peak intensity value.
The temporal structures of the wave-mixing pulses NIR$_1$ and NIR$_2$ are optimized and measured to be $\tau_{FWHM}(NIR_1) = (6.5\pm0.3)$\,fs and $\tau_{FWHM}(NIR_2) = (6.7\pm0.3)$\,fs using a home-built transient-grating FROG apparatus \cite{rupprecht2023flexible}.
The resulting NIR peak-intensity of this experimental campaign ranges from $7.0 \times 10^{10}\,$W/cm$^2$ for the 2s2p state to $1.5 \times 10^{12}\,$W/cm$^2$ for the higher lying 2s$n$p Rydberg states.

\section{\label{sec:couplings}FWM coupling pathways}
\begin{figure*}[t]
\centering
\includegraphics[width=0.85\textwidth]{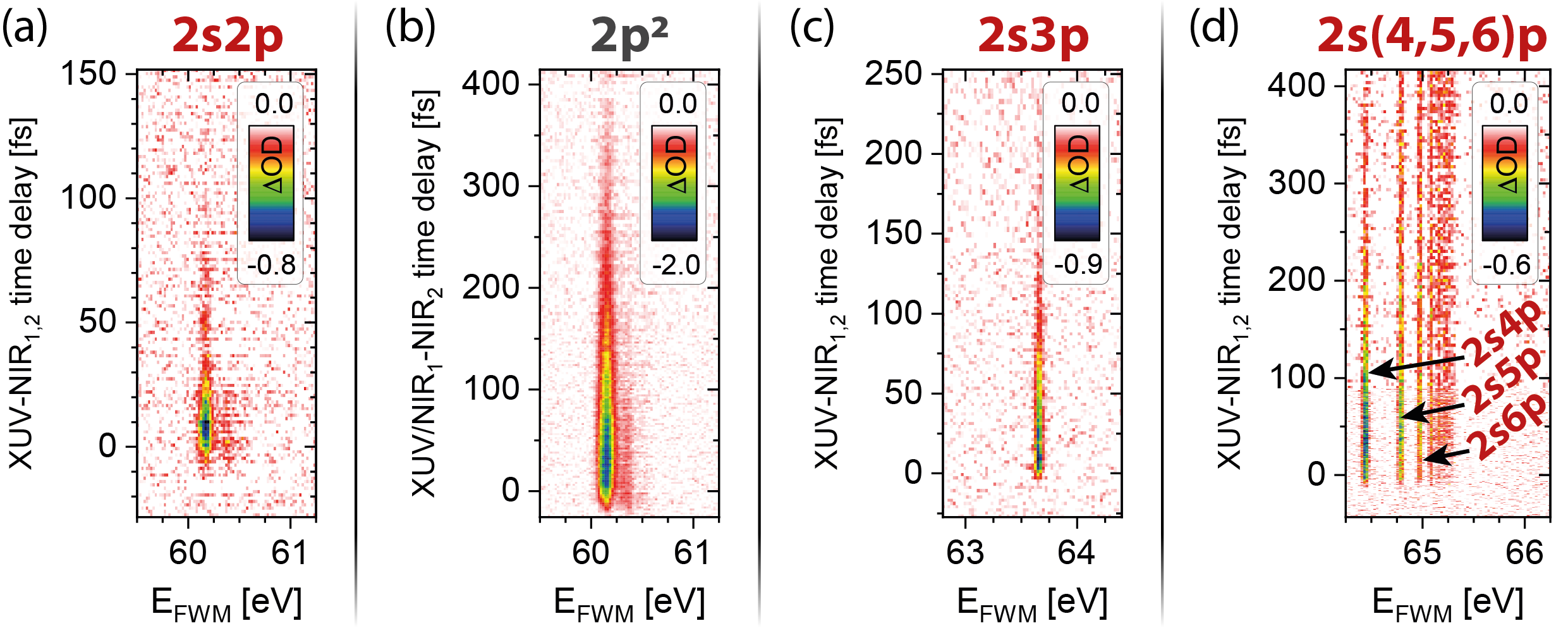}
\caption{XUV/NIR FWM time-delay measurements of doubly-excited states in helium. The respective traces show the time-dependent FWM spectra $\Delta$OD($\omega_{XUV}, t$) for (a) the 2s2p bright state [bright-state scan configuration ($t = \tau_1 = \tau_2$)] at intensity of each NIR pulses of $I_{NIR1,2} = 7.0 \times 10^{10}$\,W/cm$^2$, (b) the 2p$^2$ dark state (dark-state scan configuration with $t = \tau_2$, $\tau_1 = 0$) at $I_{NIR1,2} = 1.2 \times 10^{11}$\,W/cm$^2$, (c) the 2s3p bright state at $I_{NIR1,2} = 8.0 \times 10^{11}$\,W/cm$^2$, and (d) the 2s$n$p bright states with $n\in \{4,5,6\}$ at $I_{NIR1,2} = 1.5 \times 10^{12}$\,W/cm$^2$.}
\label{fig:2Dtraces}
\end{figure*}
In order to correlate the state-specific doubly-excited state lifetime information with a FWM signature, a quantum-path analysis as shown in Fig.~\ref{fig:fwmscheme} is helpful.
For the presented data, the lower FWM signal is recorded and hence Fig.~\ref{fig:fwmscheme} illustrates coupling schemes that are phase-matched for the lower FWM emission in Fig.~\ref{fig:expsetup}.
First, the broadband XUV pulse induces dipole transitions from the 1s$^2$ ground state to the 2s$n$p bright states.
The 2s$n$p Rydberg series converges to the ionization threshold with a remaining, excited 2s electron at E$_{IP}$($n$=2)$ \ = 65.4$\,eV.\par
In general, three different coupling scenarios are evident from Fig.~\ref{fig:fwmscheme}.
Each of these coupling scenarios represents doubly-excited state coupling schemes that lead to the lifetime extraction for at least one of the 2s$n$p states in a bright-state scan configuration.
As the pathway analysis will show, however, only one dark state lifetime (the one of the 2p$^2$ state) is directly accessible with our experimental conditions.\par
Firstly, Fig.~\ref{fig:fwmscheme}(a) illustrates the basic dark state scenario, which constitutes a resonant coupling to only one discrete dark state. 
This is the case for the $\Lambda$-coupling of the 2s2p bright state to the 2p$^2$ dark state.
In principle, a ladder coupling scheme by $-2 \times \vec{k}_{NIR_2}$ from the 2s3p state via the 2p$^2$ state to the 2s2p state also results in a FWM emission at the 2s2p resonance energy as shown by dashed arrows in Fig.~\ref{fig:fwmscheme}(a). 
However, the dipole transition moment from the 2s3p state to the 2p$^2$ state (d(2p$^2$$\leftrightarrow$2s3p)\,=\,-0.81\,a.u. \cite{ott2014reconstruction}) is significantly smaller than from the 2s2p to the 2p$^2$ (d(2s2p$\leftrightarrow$2p$^2$)\,=\,2.17\,a.u. \cite{loh2008femtosecond}) state.
Hence, the 2s3p contribution to the measured FWM signal is negligible for low NIR$_{1,2}$ intensities.\par
Secondly, Fig.~\ref{fig:fwmscheme}(b) depicts $\Lambda$- and V-type couplings from the 2s3p state to multiple discrete dark states as well as to parts of the 2s continuum above the E$_{IP}$($n$=2) threshold.
Due to the possible large manifold of dark-state couplings, the 2s3p level cannot provide selective coupling pathways to single dark states with the bandwidth of the NIR spectrum utilized in this experiment.
Such a selective coupling pathway is required to unambiguously extract lifetimes for dark states.
Therefore, the 2p$^2$ lifetime via $\Lambda$-coupling of the 2s2p state is the only dark-state lifetime accessible in this study.
Similar to Fig.~\ref{fig:fwmscheme}(a), a resonant ladder coupling is possible as well, originating from the 2s2p state via $+2 \times \vec{k}_{NIR_1}$.
A comparison of total transition dipole momenta as conducted for the FWM emission at the 2s2p resonance energy is difficult in this case due to the variety of coupling options from the 2s3p state.
Nevertheless, a potential influence of the ladder-coupling pathway on the overall FWM emission will be constrained to time delays close to the temporal overlap and will not affect the exponential fit for longer time delays as the 2s2p lifetime is significantly shorter than the one of the 2s3p state.\par
Thirdly, the 2s$n$p states with $n\geq4$ couple solely to the dark-blue shaded 2s continuum in Fig.~\ref{fig:fwmscheme}(c) via a $\Lambda$ scheme.
No resonant dark-state coupling is expected in this case as the first dark state that is located in the 2s continuum is the 3s$^2$ state at 69.4\,eV \cite{burgers1995highly}.
Hence, this resonance is beyond the maximum NIR photon energy for the bandwidth of the NIR spectrum even when coupling directly from the 2s threshold (equaling a maximum reachable resonance energy of 67.9\,eV). %  significant contribution from isolated dark-states is expected.
By minimizing the utilized NIR FWM intensity, nonresonant coupling contributions as they were observed and analyzed by Gaynor \textit{et al.} in argon \cite{gaynor2023nonresonant} in the $4\times10^{12}$\,W/cm$^2$ NIR intensity regime are avoided, which is the prerequisite for accurate natural lifetime determinations of quantum states. \par
This pathway analysis of the expected FWM coupling channels gives an intuitive idea about which state lifetimes are accessible under the utilized experimental conditions.
Moreover, the necessity of different NIR intensities for the lifetime measurement of different states is expected based upon the coupling scenarios in Fig.~\ref{fig:fwmscheme}. 
Indeed, it turns out that the optimal NIR peak intensities that allow for a measurable FWM signature while avoiding NIR-induced lifetime alterations vary for the experimental lifetime characterization of the 2s2p, the 2p$^2$, the 2s3p, and the 2s$n$p with $n\in \{4,5,6\}$ states as shown in the summary Table~\ref{tab:tableSummary}.
\begin{figure*}[t]
\centering
\includegraphics[width=\textwidth]{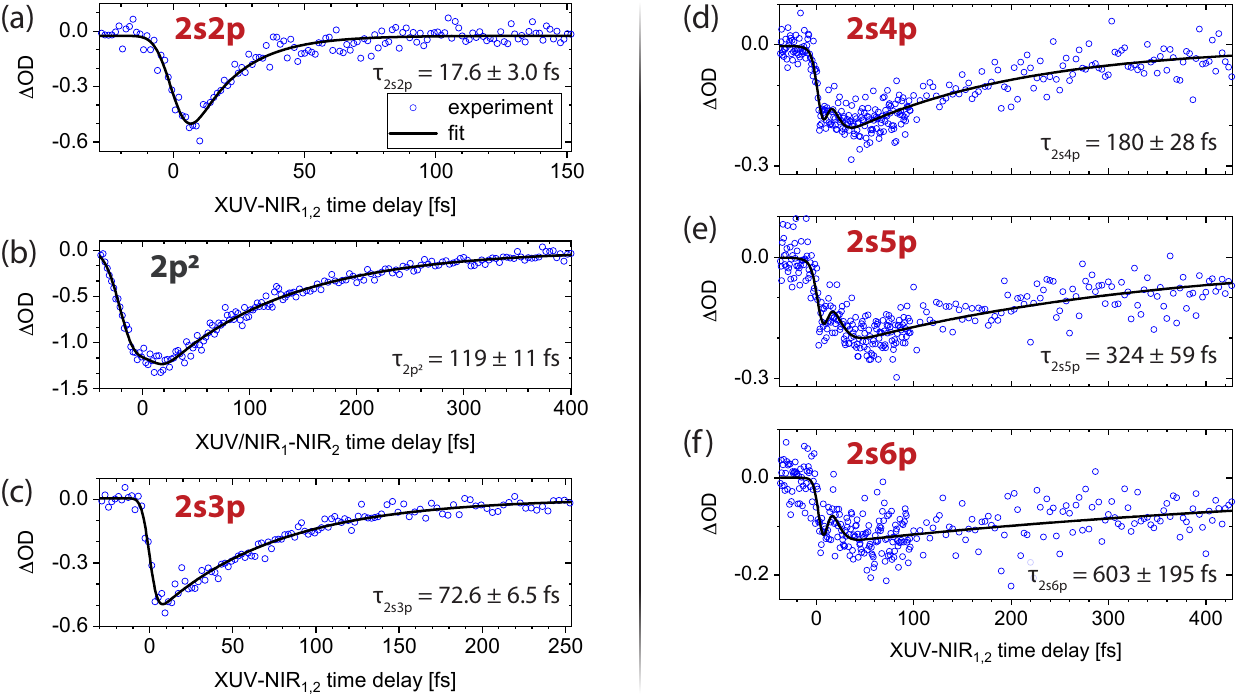}
\caption{Time dependence of FWM signatures. The blue circles represent the time-dependent experimental lineouts resulting from averaging each FWM signature of Fig.~\ref{fig:2Dtraces} over its spectral FWHM. The decay dynamics of the (a) 2s2p, (b) 2p$^2$, (c) 2s3p, (d) 2s4p, (e) 2s5p, and (f) 2s6p states are fitted (black solid lines; details in the main text) and the resulting state lifetimes given within a 95\% confidence interval.}
\label{fig:FWMtimeLineouts}
\end{figure*}

\section{\label{sec:measurements}FWM measurements and discussion}
Overall, bright-state scans at three different NIR intensities were performed to characterize the 2s2p, 2s3p, and 2s$n$p ($n\in \{4,5,6\}$) doubly-excited-state lifetimes. 
In addition, dark-state scans for characterization of the 2p$^2$ lifetime at the 2s2p FWM-emission photon energy were conducted.
The respective time-resolved traces are shown in Fig.~\ref{fig:2Dtraces}.
Here, the data are depicted in terms of differential absorbance 
\begin{equation}
    \Delta OD(\omega_{XUV}, t) = - \log_{10}\left( \frac{I_{FWM}(\omega_{XUV}, t)}{I_{XUV,ref}(\omega_{XUV}, t)}\right) \ ,
\end{equation}
with the time-dependent FWM spectral intensity as $I_{FWM}(\omega_{XUV}, t)$, the reference XUV intensity with blocked NIR wave-mixing beams as $I_{XUV,ref}(\omega_{XUV}, t)$ and the measurement time-delay as $t$.
In the case of a bright-state scan, $t$ is equal to $\tau_1$ and $\tau_2$ as both NIR beams are time-delayed with respect to the incoming XUV pulse.
For dark-state scans, NIR$_1$ remains in temporal overlap with the XUV ($\tau_1 = 0$) while $\tau_2$ is scanned ($t = \tau_2$).
While the background-free character of the noncollinear FWM should theoretically cause $I_{XUV,ref}(\omega_{XUV}, t)$ to completely vanish, in reality the remaining stray light and CCD read-out noise are included in this term.
Hence, the representation of the FWM signal in terms of $\Delta$OD accounts for these inevitable experimental noise sources and can be therefore viewed as a signal-to-noise quantification of the FWM measurement.\par

\begin{table*}
\caption{Summary of measured doubly-excited state lifetimes near the He$^+$ (n=2) ionization threshold. The lifetimes, which were measured using FWM spectroscopy, are given within a 95\% confidence interval. In addition, the NIR FWM peak intensity of the respective measurement is noted. Furthermore, experimental and theoretical spectral linewidths from literature are given as well as the deduced lifetimes from the respective linewidth values.}
\begin{ruledtabular}
\begin{tabular}{cccccccccc}
 \begin{tabular}{@{}c@{}}electronic\\state\end{tabular}& \begin{tabular}{@{}c@{}}state \\ character\end{tabular}&\begin{tabular}{@{}c@{}}resonance\\ energy [eV]\end{tabular}&\begin{tabular}{@{}c@{}}exp. line-\\ width [meV]\end{tabular}&\begin{tabular}{@{}c@{}}exp. width\\lifetime [fs]\end{tabular}&\begin{tabular}{@{}c@{}}theo. line-\\ width [meV]\end{tabular}&\begin{tabular}{@{}c@{}}theo. width\\lifetime [fs]\end{tabular}&\begin{tabular}{@{}c@{}}FWM measured\\ lifetime [fs]\end{tabular}&\begin{tabular}{@{}c@{}}NIR peak\\ int. [W/cm$^2$]\end{tabular}\\ \hline
 \rule{0pt}{3ex}%  EXTRA vertical height
 $^1$P$^o$ 2s2p&bright&60.15\footnotemark[1]&$37\pm1$\footnotemark[1]&$17.8^{+0.5}_{-0.5}$&37.37\footnotemark[2] &17.6&$17.6\pm3.0$&$7.0 \times 10^{10}$\\
 \rule{0pt}{3ex}%  EXTRA vertical height
 $^1$S$^e$ 2p$^2$&dark&62.09\footnotemark[3]& $7\pm 7$\footnotemark[3]&$94^{+\infty}_{-47}$&5.87\footnotemark[4]&112.1&$119\pm11$&$1.2 \times 10^{11}$\\
 \rule{0pt}{3ex}%  EXTRA vertical height
 $^1$P$^o$ 2s3p&bright&63.66\footnotemark[1]&$10\pm1$\footnotemark[1]&$66^{+7}_{-6}$&8.20\footnotemark[2]&80.3&$72.6\pm6.5$&$8.0 \times 10^{11}$\\
 \rule{0pt}{3ex}%  EXTRA vertical height
 $^1$P$^o$ 2s4p&bright&64.47\footnotemark[1]&$4\pm 0.5$\footnotemark[1]&$165^{+24}_{-18}$&3.49\footnotemark[2]&188.6&$180\pm28$&$1.5 \times 10^{12}$\\
 \rule{0pt}{3ex}%  EXTRA vertical height
 $^1$P$^o$ 2s5p&bright&64.82\footnotemark[1]&$2\pm0.3$\footnotemark[1]&$329^{+58}_{-43}$&1.79\footnotemark[2]&367.7&$324\pm59$&$1.5 \times 10^{12}$\\
 \rule{0pt}{3ex}%  EXTRA vertical height
 $^1$P$^o$ 2s6p&bright&65.01\footnotemark[2]&&&1.03\footnotemark[2]&639.0&$603\pm195$&$1.5 \times 10^{12}$\\
\end{tabular}
\end{ruledtabular}
\footnotetext[1]{Reference \cite{domke1996high}.}
\footnotetext[2]{Reference \cite{rost1997resonance}.}
\footnotetext[3]{Reference \cite{deharak2006ejected}.}
\footnotetext[4]{Reference  \cite{burgers1995highly}.}
\label{tab:tableSummary}
\end{table*}

\subsection{\label{subsec:lifetimes}Lifetime extraction}

In order to extract the respective quantum-state lifetimes from the time-delay measurement traces of Fig.~\ref{fig:2Dtraces}, slices along the spectral axis with a width of approximately the full-width-at-half-maximum (FWHM) of the FWM spectral signatures are averaged.
The respective time-dependent FWM lineouts are shown in Fig.~\ref{fig:FWMtimeLineouts}.
The FWM signals are fitted (black solid lines in Fig.~\ref{fig:FWMtimeLineouts}) via the convolution of a Gaussian function with a monoexponential decay. 
As described in the theoretical FWM paper on doubly-excited states in helium by Mi \textit{et al.} \cite{mi2021method}, the lifetimes of some FWM signatures require a superposition of the Gaussian/exponential convolution with an additional Gaussian function to account for nonresonant pathways contributing to the FWM signals at the overlap of all three pulses. 
This additional Gaussian transient-grating-like signature hence encodes information about the NIR pulse durations and temporal overlaps instead of state lifetimes.
In the data presented here, a fit model that includes an additional Gaussian enables a precise fit of the 2p$^2$ and the 2s(4,5,6)p FWM signatures presented in Fig.~\ref{fig:FWMtimeLineouts}(b), and (d)-(f), respectively.\par
\begin{figure*}[t]
\centering
\includegraphics[width=\textwidth]{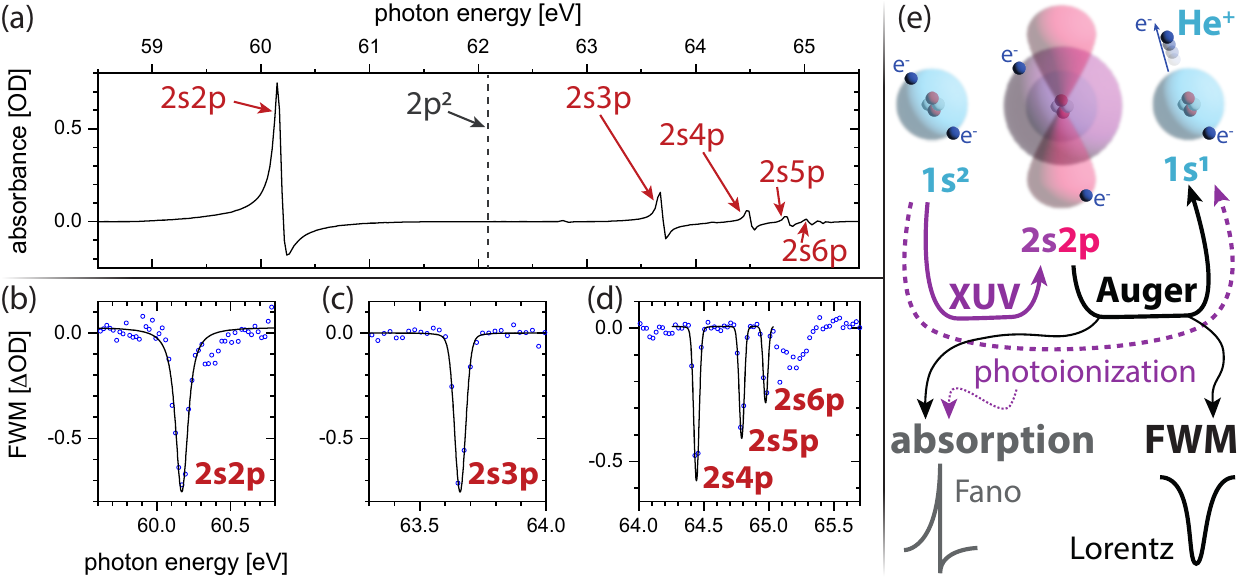}
\caption{FWM spectral line shapes in comparison to absorption line shapes. (a) Static XUV absorbance spectrum with the annotated bright states in red and the 2p$^2$ dark state in gray. FWM spectra (binned around max signal -- blue circles -- and Voigt fits as black solid lines) of the (b) 2s2p, (c) 2s3p, and (d) 2s$n$p with $n \in \{4,5,6\}$. (e) Excitation/relaxation pathways contributing to the absorption and the FWM signals on the example of the 2s2p state. In absorption spectroscopy, the direct photoionization channel (purple dashed arrow) interferes with the 2s2p decay mechanism dominated by Auger decay. This interference gives rise to a Fano line shape. In the noncollinear FWM case, however, the Auger channel is not heterodyned with the photoionization channel. Hence, the FWM signals reflects the natural lifetimes of the state as a Lorentzian peak.}
\label{fig:FWMlineshapes}
\end{figure*}
The resulting fitted lifetime values within a 95\% confidence interval are given in the lineout plots in Fig.~\ref{fig:FWMtimeLineouts} and compared to lifetimes deduced from experimentally measured and theoretically calculated spectral linewidths in Table~\ref{tab:tableSummary}. 
This comparison shows excellent agreement between the experiment and the published theoretical work with all measured lifetimes matching the theoretically deduced ones within the uncertainty intervals except for the 2s3p lifetime, where the theoretical value is just slightly above the experimental confidence interval. 
Table~I also summarizes experimentally measured spectral linewidth values from the literature. 
The uncertainties of these high-precision spectral measurements at synchrotrons are 0.3\,meV or more. 
For the 2p$^2$ dark state, no static XUV synchrotron-based high-resolution spectrum is available due to the 1s$^2\rightarrow2p^2$ transition being dipole forbidden. 
Electron impact studies, on the other hand, have limited resolution.
In the case of the 2p$^2$ state the experimental spectral linewidth uncertainty is on the order of the resonance width itself \cite{deharak2006ejected}.
Due to the experimental time-delay range of around 0.5\,ps and the trend towards smaller measured FWM-signal intensities for higher-lying Rydberg states, the 2s6p lifetime measurement ($\tau_{2s6p} = 603 \pm 195$\,fs) marks the threshold for a meaningful lifetime extraction in the presented measurement. \par
These aspects highlight the power of noncollinear XUV/NIR FWM spectroscopy to characterize bright and dark states directly in the time domain without being constrained by the resolution limitations of spectral measurements. 
Furthermore, the doubly-excited states in helium are a remarkable showcase scenario for these capabilities as highly accurate \textit{ab-initio} calculations are available for these resonances with few tens to hundreds of femtosecond lifetimes \cite{rost1997resonance, burgers1995highly}. 

\subsection{\label{subsec:decay_mechanism}Interpretation of decay mechanisms}

The decay mechanisms leading to lifetimes of doubly excited states in helium have been previously discussed \cite{odling2000radiative,liu2001radiative, lambourne2003experimental}.
In general, the quantum-state lifetime is given as a linear combination of the radiative decay and Auger-Meitner decay. 
While the radiative $2p \rightarrow 1s$ decay in He$^+$ is 0.1\,ns \cite{drake1992he+}, Auger-Meitner decay dominates the overall decay of the 2s$n$p Rydberg series up $n \approx 30$ \cite{liu2001radiative}. 
The Auger-Meitner decay follows the $n^{*3}$ law \cite{fano1965line} with $n^{*}$ being the effective principle quantum number, which has been systematically calculated \cite{burgers1995highly,rost1997resonance}.\par
Strikingly, the $^1$S$^e$ 2p$^2$ dark state has a six times longer lifetime than the $^1$P$^o$ 2s2p state. 
This distinct difference in lifetimes is explained by the decay pathways given by propensity rules for autoionization \cite{rost1990propensity}. 
Amongst different theoretical concepts and approximations (a thorough review of theory efforts to describe doubly-excited states in helium can be found in Ref. \cite{tanner2000theory}), the so-called "molecular approximation" \cite{feagin1986molecular,feagin1988molecular} grants an intuitive access to understanding the different decay mechanisms according to propensity rules.
In the molecular approximation, the two electrons and the nucleus of the helium atom are treated in analogy to the two nuclei and the one electron of the H$_2^+$ molecule by reversing the roles of the electrons and nuclei.
Hence, the inter-nuclear axis of the H$_2^+$ molecules becomes the inter-electronic axis in the helium atom. 
This idea leads to potential energy curves depending on the distance between the electrons similar to the potential energy curves along the inter-nuclear axis known from molecular dynamics.
The resulting set of quantum numbers completely describes doubly excited states of helium and the nodal structure of the two-electron wavefunction, which in turn manifests propensity rules for the decay of the respective state. 
This nodal structure in the two-electron wavefunction gives rise to so-called avoided crossings in the adiabatic potential curves within the molecular approximation of helium. 
Non-adiabatic transitions of these avoided crossings, in turn, correspond to autoionization \cite{rost1991saddle, rost1990propensity}.\par
The considerable difference in the 2s2p versus the 2p$^2$ lifetime is explained by the different propensity-rule mechanism for the non-adiabatic coupling of these states in the molecular approximation picture.
While the autoionization mechanism of the 2s2p state relies on rotational coupling, this mechanism is not allowed for the 2p$^2$ state and it undergoes strongly suppressed relaxation that results in a change of the number of wavefunction nodes between the inner and the distant outer electron (change in quantum number $n_{\lambda}$ in prolate spheroidal coordinates) \cite{rost1990propensity,rost1997resonance}.
As a result, the 2p$^2$ dark-state lifetime is calculated to be about six times longer compared to the 2s2p lifetime, which is experimentally verified by the here presented FWM measurements (see Table \ref{tab:tableSummary}).

\subsection{\label{subsec:lineshapes}Spectral line-shape analysis}

The doubly-excited state character of the absorption peaks comprising the 2s$n$p Rydberg series has consequences concerning their spectral line shapes. 
While the radiative dipole decay of an excited state results in a Lorentzian spectral line shape, the doubly-excited states have Fano line shapes as depicted in the static absorbance spectrum in Fig.~\ref{fig:FWMlineshapes}(a). 
This is caused by the interference of the direct photoionization channel [He(1s$^2$) + $\gamma_{XUV} \rightarrow$\,He$^+$(1s$^1) + e^-$] induced by an XUV photon $\gamma_{XUV}$ with the Auger-Meitner decay channel via the doubly excited states [He(1s$^2$) + $\gamma_{XUV} \rightarrow$\, He(2s$n$p)\,$\rightarrow$\,He$^+$(1s$^1) + e^-$]. 
These two ionization pathways are illustrated using the 2s2p state as an example in Fig.~\ref{fig:FWMlineshapes}(e).
The interference of these two pathways represents a quantum interferometer that gives rise to the Fano line shapes in the static absorbance spectrum \cite{fano1961effects}. \par
In contrast, each background-free FWM measurement presented in Fig.~\ref{fig:2Dtraces}(a), (c), and (d) is spectrally well described by a Voigt profile: the convolution of the natural Lorentzian linewidth with a Gaussian response function of the XUV spectrograph. 
The spectral lineouts at temporal overlap are shown in Fig.~\ref{fig:FWMlineshapes}(b)-(d) including the respective Voigt fits.
Here, the fits utilize the theoretically calculated linewidths $\Gamma$ from Table~I as Lorentzian widths and result in consistent Gaussian contributions translating to a spectrograph resolution of E$_{XUV}$/$\Delta$E$_{XUV}$\,= 1400 for the 2s3p to 2s6p signatures in Fig.~\ref{fig:FWMlineshapes}(c) and (d). 
This is in good agreement with a spectrograph resolution of E$_{XUV}$/$\Delta$E$_{XUV}$\,= 1300 extracted from fitting the Fano line shapes in the static absorbance spectrum shown in Fig.~\ref{fig:FWMlineshapes}(a) with the known resonance parameters. \par
For the 2s2p FWM spectrum in Fig.~\ref{fig:FWMlineshapes}(b), however, the extracted linewidth utilizing a fixed resolution of E$_{XUV}$/$\Delta$E$_{XUV}$\,= 1400 results in a Lorentzian width that is considerably larger than the expected 37.4\,meV. 
Furthermore, a spectral feature begins to appear around 60.35\,eV in time overlap.
While the fitted lifetime value of $(17.6 \pm 3.0)$\,fs is in excellent agreement with the theoretically predicted one, the distortions of the spectral width and line shape are indications of the onset of strong-field effects. 
The impact of such strong-field effects on FWM signatures will be analyzed in detail in a future publication \cite{rupprecht2024rabi}. 
Therefore, the NIR intensities utilized here in the 2s2p bright-state scan [Fig.~\ref{fig:2Dtraces}(a)] are considered to be at the threshold where the FWM NIR pulses themselves start to impact an accurate lifetime measurement. 
This is further discussed in Section \ref{sec:rabi}.\par
Overall, the extracted FWM line shapes exhibit the Lorentzian emission spectral profiles of the natural lifetime decays.
This clear difference in spectral line shapes between the absorption and the FWM signatures can be explained by the phase-matched, background-free nature of the noncollinear FWM experiment. 
While the 2s$n$p decays, which are probed by the FWM pathways described in Fig.~\ref{fig:fwmscheme}, are phase-matched according to Eq.~\ref{eq:phase-matching}, direct photoionization does not contribute to the off-axis FWM signals for the measurements in Fig.~\ref{fig:2Dtraces}.
Due to the moderate NIR intensities, no broadband FWM emission is observed that would originate from a nonresonant coupling of the 1s$^2$ ground state to the continuum and could interfere with the doubly-excited state pathways of Fig.~\ref{fig:fwmscheme}.

\subsection{\label{sec:rabi}Onset of strong-field effects}
As the 2s2p bright-state scan depicted in Fig.~\ref{fig:2Dtraces}(a) and the respective spectral lineout in Fig.~\ref{fig:FWMlineshapes}(b) show the onset of spectral distortions in the FWM signal (the spectral line shape is broadened and a spectral feature appears around 60.35\,eV), the NIR intensity at which this scan was taken ($7.0\times10^{10}\,$W/cm$^2$) can be viewed as the intensity threshold for accurate lifetime extraction via XUV/NIR FWM spectroscopy.
NIR intensities above this threshold result in strong-field effects altering the measured state lifetime as will be shown in a future publication \cite{rupprecht2024rabi}.
The 2s2p bright-state case is ideally suited to calculate an approximate value for the onset of strong-field effects as only one channel, the 2s2p\,$\leftrightarrow$\,2p$^2$ transition, dominates the FWM signal [Fig.~\ref{fig:fwmscheme}(a)].
Furthermore, the transition dipole moment of this transition is known and given by d(2s2p$\leftrightarrow$2p$^2$)\,=\,2.17\,a.u. \cite{loh2008femtosecond}.
With the total NIR peak intensity of the bright-state scan with both NIR pulses in temporal overlap (I$_{peak} = 2\times70$\,GW/cm$^2$), the maximum Rabi frequency is $\Omega_{max} = 179$\,THz, corresponding to a Rabi-cycle period of T$_{Rabi} = 2 \pi  / \Omega_{max} = 35$\,fs.
The NIR electric field envelope contains a pulse area of approximately A$_{pulse} = 10$\,fs\,$\times$\,I$_{peak}$.
Therefore, as a rule-of-thumb value, the combined NIR-pulse intensity in the FWM scheme should remain below a value leading to about 0.3 Rabi cycles for bright-state scans to guarantee an accurate quantum-state lifetime determination of the helium doubly-excited states.
This estimation can be made for any other quantum system with consideration of the appropriate transition dipoles to prevent Rabi-flopping dynamics from interfering with the natural decay.

\section{\label{sec:outlook}Summary and outlook}

In summary, the 2s$n$p ($n \in \{2,3,4,5,6\}$) as well as 2p$^2$ doubly-excited state lifetimes are characterized directly in the time domain utilizing noncollinear XUV/NIR FWM spectroscopy.
The extracted lifetimes are in excellent agreement with theoretical predictions and showcase the ability of noncollinear FWM spectroscopy to accurately target lifetimes from tens to hundreds of femtoseconds and beyond in a quantum-state specific manner. 
Furthermore, the spectral line shapes of the measured FWM signals are of Lorentzian nature independent of a more complex Fano absorption line shape. 
This is a direct consequence of the phase-matching condition of noncollinear FWM, which separates the decay of highly excited states from different potentially interfering mechanisms like photoionization. 
As the NIR intensity in the FWM scheme is an important parameter with its trade-off between FWM signal intensity versus influence on the quantum states themselves, an upper threshold of 0.3 Rabi cycles for an accurate lifetime determination is identified. \\
\indent In principle, the range of lifetimes that can be characterized using XUV/NIR FWM spectroscopy could be easily expanded to the picosecond regime by introducing longer optical delay lines. 
This could be of interest for bridging the gap to time-resolved XUV fluorescence spectroscopy as a lifetime determination technique \cite{lambourne2003experimental} for longer-lived doubly-excited states in helium.
On the other extreme, quantum state lifetimes of 7\,fs and below have been characterized with XUV/NIR FWM spectroscopy \cite{lin2021coupled,puskar2023measuring}.
With an active time-delay stabilization scheme \cite{sabbar2014combining,schlaepfer2019phase} and sub-cycle light transients \cite{wirth2011synthesized}, precise few-femtosecond core-hole lifetime \cite{haynes2021clocking} extraction will be feasible with a table-top FWM setup.
Furthermore, state-of-the-art high repetition rate HHG sources \cite{harth2018compact,fattahi2014third} will enhance the FWM signal-to-noise ratio by at least one order of magnitude.
Therefore, even lower FWM NIR intensities can be utilized for quantum-state lifetime characterization in order to avoid strong-field effects while still maintaining a measurable FWM signal intensity.
Employing high FWM NIR intensities, on the other hand, can probe Rabi-cycling dynamics of dark-states directly in the time domain with an all-optical method as will be reported elsewhere \cite{rupprecht2024rabi}.
The multi-pulse scheme employed in FWM spectroscopy can be also utilized to study and manipulate electronic exchange effects \cite{rupprecht2022laser} in atoms and molecules.
Overall, this study contributes to a better understanding of the experimental parameters needed for future FWM measurements of quantum-dynamical processes, which can possibly be applied to soft x-ray excited states in organic polyatomic molecules \cite{barreau2020efficient,ren2018attosecond} in the future.

\begin{acknowledgments}
This work was performed by personnel and equipment supported by the Office of Science, Office of Basic Energy Sciences through the Atomic, Molecular, and Optical Sciences Program of the Division of Chemical Sciences, Geosciences, and Biosciences of the U.S. Department of Energy (DOE) at Lawrence Berkeley National Laboratory under Contract No. DE-AC02-05CH11231. 
P.~R. acknowledges funding by the Alexander von Humboldt Foundation (Feodor-Lynen Fellowship).
N.~G.~P. acknowledges funding from Soroptimist International of the Americas (Founder Region Fellowship).
\end{acknowledgments}

\bibliography{apssamp}% Produces the bibliography via BibTeX.

\end{document}